\documentclass[cameraready]{Interspeech}

\title{ANCHOR: Autoregressive Non-intrusive Chunk-Ordered Refinement for Joint Multi-Resolution Speech Quality Modeling}

\author[affiliation={1}, orcid=0009-0000-6819-0195]{Zhuoyan}{Tao}
\author[affiliation={2}, orcid=0000-0002-9050-8304]{Jiatong}{Shi}
\author[affiliation={2}, orcid=0000-0002-5990-363X]{Hye-jin}{Shim}
\author[affiliation={2}, orcid=0000-0002-5970-8631]{Shinji}{Watanabe}

\address{
    $^1$ University of Southern California, USA \\
    $^2$ Carnegie Mellon University, USA
}

\email{terryt@usc.edu, jiatongs@cs.cmu.edu, shimhz6.6@gmail.com, shinjiw@ieee.org}


\usepackage{comment}
\usepackage[utf8]{inputenc}
\usepackage{stfloats}
\usepackage{placeins}
\usepackage{enumitem} 
\usepackage{cite}
\setlist[itemize]{noitemsep, topsep=0pt, leftmargin=*}
\usepackage[activate={true,nocompatibility},tracking=true,kerning=true,spacing=true,stretch=10,shrink=10]{microtype}

\begin{document}

\maketitle

\begin{abstract}
While speech quality is typically assessed on complete utterances, streaming and generative systems require incremental estimation from partial audio. Existing predictors assume full context, degrading on prefix-constrained inputs. Extending ARECHO, we propose ANCHOR, reformulating incremental assessment as a multi-resolution autoregressive task. It models chunk- and utterance-level quality within a single decoder using dual-resolution tokens and a resolution-aware hierarchy for coarse-to-fine refinement. Experiments show substantial robustness under partial input, including a 48\% PLCMOS error reduction on 2-second prefixes. Convergence analysis reveals a 4–6\,s effective perceptual context horizon. A stress test further isolates structured extrapolation biases under localized corruption. Results demonstrate that hierarchical supervision improves incremental prediction and elucidates how perceptual quality accumulates over time.
\end{abstract}

\section{Introduction}

Many practical speech systems, such as streaming communication~\cite{shi2021emformer}, speech enhancement~\cite{defossez2020realtime}, and autoregressive generative speech models~\cite{wang2017tacotron, kalchbrenner2018efficient, borsos2023audiolm, kharitonov2023spear, chen2025valle, le2023voicebox, wang2025sparktts}, operate under partial acoustic context, motivating perceptual quality estimation before full utterances are observed. This incremental setting mirrors human speech perception, where listeners process acoustic signals as they unfold in time rather than waiting for utterance completion~\cite{MarslenWilson1987FunctionalParallelism}. In contrast, most existing objective and learned quality predictors, including PESQ~\cite{rix2001pesq} and ViSQOL~\cite{hines2015visqol}, are designed under a full-context assumption: a complete utterance is available before a score is produced.

Beyond intrusive metrics, many learned non-intrusive quality predictors are likewise formulated as utterance-level regression problems, where a single global score is inferred from representations aggregated over the entire signal. While such full-context aggregation promotes stability under complete input, it implicitly assumes that all acoustic information is available at inference time and that perceptual quality is inherently a holistic property.

This limitation is particularly pronounced for localized artifacts such as short packet-loss bursts, clipping events, or brief background intrusions. Because these distortions are temporally sparse, prediction mechanisms that rely on utterance-level pooling or global contextual integration may attenuate their influence until sufficient future context is observed. As a result, early quality estimates may not faithfully reflect perceptual degradation.

To address the mismatch between full-context inference and prefix-constrained evaluation, we reformulate quality estimation as a multi-resolution autoregressive prediction problem built on ARECHO~\cite{shi2025arecho}. The model jointly predicts chunk-level and full-utterance perceptual metrics within a unified decoding framework. To stabilize supervision across resolutions, we introduce a \textbf{resolution-aware decoding hierarchy}, in which chunk-level metric tokens are generated first, followed by full-utterance tokens. This coarse-to-fine schedule mitigates supervision conflict between local and global objectives while preserving a shared encoder–decoder architecture.

We note that our evaluation targets operational pseudo-MOS estimators
(PLCMOS, UTMOS, NISQA) that serve as standard metrics in production
systems (e.g., Microsoft Teams) and community benchmarks (e.g., DNS
Challenge, VoiceMOS Challenge). Our contribution is not a new perceptual
ground truth but rather a new \emph{inference regime}: predicting these
established metrics incrementally from partial input, which none of them
natively support. Our contributions are as follows:
\begin{itemize}
    \item We introduce joint chunk-level and full-utterance multi-metric supervision within a unified autoregressive framework.
    \item We propose a resolution-aware decoding hierarchy that enforces structured coarse-to-fine prediction under prefix-constrained inference.
    \item Through prefix-to-full convergence analysis, we show that chunk-first supervision improves packet-loss–sensitive modeling under partial context and reveals an effective perceptual context horizon for incremental quality estimation.
    \item We design a controlled distortion stress test to isolate partial-context behavior, demonstrating that ANCHOR maintains stable extrapolation under localized corruption. This confirms the model captures genuine perceptual accumulation rather than merely overfitting to artificial truncation boundaries.
\end{itemize}
\section{Related Work}

Traditional speech quality assessment has relied on intrusive metrics such as PESQ~\cite{rix2001pesq}, ViSQOL~\cite{hines2015visqol}, and STOI~\cite{taal2011stoi}. While accurate, these methods require a clean, time-aligned reference signal, limiting their utility in real-world scenarios where only the degraded signal is available. To address this, non-intrusive (no-reference) assessment has emerged as a critical field. Early deep learning approaches, including UTMOS~\cite{saeki2022utmos},  DNSMOS~\cite{reddy2021dnsmos}, and MOSNet~\cite{lo2019mosnet}, framed this as a scalar regression task targeting Mean Opinion Scores (MOS). Other frameworks, such as NORESQA~\cite{manocha2021noresqa}, NISQA~\cite{mittag2021nisqa}, NOMAD~\cite{nomad2024} explicitly model specific signal distortions to improve interpretability and robustness.

Recent literature has expanded the scope of non-intrusive assessment beyond speech to music and sound evaluation through frameworks like SongEval~\cite{lee2023songeval} and Audiobox~\cite{tjandra2025audiobox}. Furthermore, the field has seen a shift toward more complex modeling paradigms. UrgentMOS~\cite{zhang2024urgentmos} introduced preference-based learning to better capture human subjectivity, while UniVERSA~\cite{shi2025universa} and ARECHO~\cite{shi2025arecho} leverage joint multi-metric regression and autoregressive token-based formulations to capture multi-dimensional perceptual attributes. 
While these models primarily operate on a full-utterance basis, often overlooking the temporal dynamics of quality within a single stream, there is an increasing demand for short-segment and streaming assessment.

Initial efforts in this space, such as ChunkMOS~\cite{kuhlmann2025towards}, evaluate independent segments, while ChunkSSL~\cite{tang2025chunkssl} analyzes how self-supervised learning (SSL) encoders degrade under partial input. A fundamental challenge here is the local-global coupling issue; layer-wise analyses~\cite{pasad2021layerwise} reveal that SSL representations become increasingly invariant to local signal properties at deeper layers, prioritizing global context. This creates a bottleneck for incremental prediction, as global context mixing in self-attention mechanisms can complicate prefix-based estimations~\cite{kuhlmann2025towards}.

To overcome this issue, we propose ANCHOR, a unified framework designed for hierarchical resolution-aware decoding. ANCHOR reformulates quality assessment as a structured multi-resolution autoregressive task, using ARECHO as a backbone model. By modeling both chunk-level and utterance-level perceptual quality within a single decoder, we maintain full decoder depth while mitigating the local-global coupling issue. Drawing inspiration from streaming transformer architectures~\cite{moritz2020streaming}, ANCHOR manages temporal dependencies through structured decoding order, providing a foundation for incremental quality assessment. 

\section{Method and Task Formulation}
\label{sec:method}

\subsection{Multi-Resolution Autoregressive Modeling}

ANCHOR extends ARECHO~\cite{shi2025arecho} by introducing dual-resolution metric query tokens and a resolution-aware decoding hierarchy. We leverage two properties fundamental to ARECHO: (1) a unified token space for heterogeneous perceptual metrics and (2) autoregressive conditioning that captures inter-metric dependencies. These enable hierarchical supervision, where chunk-level predictions provide a conditional anchor for full-utterance refinement within a single decoding sequence.

We formulate quality estimation as conditional token generation over a discretized metric space $\mathcal{V}$. Let $\mathcal{X}$ denote the continuous acoustic feature space. Given partial input $x_{1:t} \in \mathcal{X}^t$ and metric query tokens $\{q_1, \dots, q_K\}$, the decoder models:

\begin{equation}
P_\theta(Y \mid x_{1:t}) 
= 
\prod_{k=1}^{K} 
P_\theta(y_k \mid y_{<k}, x_{1:t}; q_k),
\end{equation}
where $q_k$ is a learned task-specific query embedding and 
$y_k \in \mathcal{V}$ is a quantized score token.

A query sequence may take the form
$\{\texttt{<UTMOS>}, \texttt{<UTMOS\_full>}, \texttt{<PLCMOS>}, \texttt{<PLCMOS\_full>}\}$.
Here, \texttt{<UTMOS>} denotes prediction on the available prefix $x_{1:t}$, whereas \texttt{<UTMOS\_full>} denotes prediction of the full-utterance UTMOS score using only $x_{1:t}$ as input. The distinction lies in the supervision target rather than the metric definition: \texttt{<UTMOS\_full>} is supervised with the full-utterance score while the input remains the prefix $x_{1:t}$.

We define chunk-level metrics as evaluations on the currently available prefix $x_{1:t}$ and full-utterance metrics as evaluations on the complete utterance $x_{1:T}$. Let $\mathcal{M}_c$ and $\mathcal{M}_f$ denote the disjoint sets of chunk-level and full-utterance metrics, respectively, with $\mathcal{M} = \mathcal{M}_c \cup \mathcal{M}_f$. For $m_c \in \mathcal{M}_c$, the target is
\[
y^{c}(t) = m_c(x_{1:t}),
\]
while for $m_f \in \mathcal{M}_f$, the target is
\[
y^{f} = m_f(x_{1:T}),
\]
although the model input remains $x_{1:t}$.

\subsection{Resolution-Aware Decoding Order}

To structure the dependency between local and global quality, we impose a specific decoding order. Let $Y^{c} \in \mathcal{V}^M$ and $Y^{f} \in \mathcal{V}^N$
denote the chunk-level and full-utterance token subsequences,
respectively, with $(M+N=K)$.
By enforcing chunk-first decoding, where $Y = [Y^{c}, Y^{f}]$, the joint probability factorizes as:

\begin{equation}
P_\theta(Y \mid x_{1:t})
=
P_\theta(Y^{c} \mid x_{1:t})
P_\theta(Y^{f} \mid x_{1:t}, Y^{c}).
\end{equation}

By avoiding the typical multi-task training pitfall where local and global predictions are interleaved without causal ordering, our hierarchy enforces a structured gradient conditioning that respects the causal relationship between local and global resolutions. By enforcing a chunk-first sequence, full-utterance prediction is explicitly conditioned on chunk-level outputs, effectively treating these localized quality estimates as intermediate latent variables. 

\begin{figure}[t]
  \centering
  \includegraphics[width=0.8\linewidth]{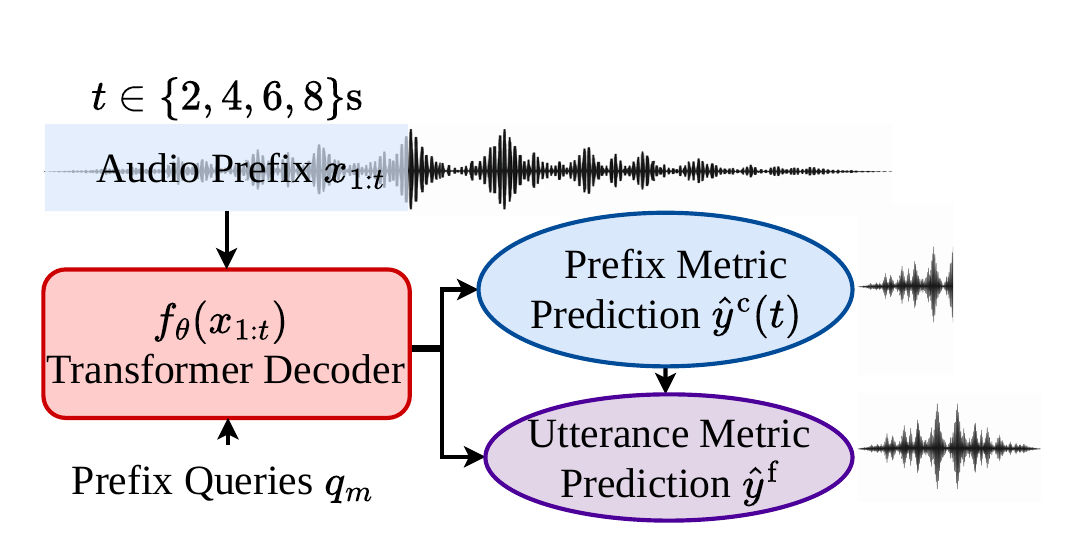}
  \caption{
  Joint multi-resolution autoregressive modeling.
  Given prefix $x_{1:t}$,
  the shared decoder generates
  chunk-level quality followed by full-utterance quality.
  }
  \label{fig:task_formulation}
\end{figure}

The model predicts $f_\theta(x_{1:t}) \rightarrow (\hat{y}^{c}(t), \hat{y}^{f})$
and is trained with paired supervision $(x_{1:t}, y^{c}(t), y^{f})$.
Figure~\ref{fig:task_formulation} illustrates the setup.

\vspace{-0.5em}
\vspace{-0.1em}

\subsection{Discrete Autoregressive Objective}
\label{sec:training_objective}

Each continuous metric value is discretized into percentile-based bins ($B{=}500$) and mapped to a token in $\mathcal{V}$, ensuring balanced token frequency; categorical metrics are mapped directly to discrete tokens without quantization. At inference, predicted tokens are mapped back via bin centroids. For heavy-tailed metrics (e.g., SI-SNR), signed log compression is applied prior to binning. The model is optimized via cross-entropy under teacher forcing:

\begin{equation}
\mathcal{L}(\theta) = \mathbb{E}_{(x, Y) \sim \mathcal{D}} \left[ - \sum_{k=1}^{K} \log P_\theta(y_k \mid y_{<k}, x_{1:t}; q_k) \right].
\end{equation}

\section{Experimental Setup}

\subsection{Dataset and Prefix Construction}


\textbf{Dataset.}
Experiments utilize the \textit{Overall Base} configuration from~\cite{shi2025arecho}, spanning 308.8 hours of clean, corrupted, and synthesized speech. The training set contains 170,013 utterances spanning
clean speech (OWSM-V3~\cite{bu2017aishell,kraaij2005ami,ardila2020commonvoice}), simulated/enhanced speech (URGENT2024~\cite{zhang2024urgent}, VoiceBank+DEMAND~\cite{veaux2013voicebank,thiemann2013demand}), and synthesized speech (VoiceMOS 2022~\cite{huang2022voicemos}, NISQA~\cite{mittag2021nisqa}). We follow the same data splits and preprocessing protocol as ARECHO~\cite{shi2025arecho}.

\textbf{Prefix Expansion.}
To realize the proposed chunk evaluation, we employ cumulative prefixes at $\{2,4,6,8\}$\,s. Utterances shorter than 2\,s are discarded for training. For utterances with duration $2 \leq T < 8$, only valid prefixes satisfying $t \leq T$ are generated.
Each prefix $x_{1:t}$ is supervised with two targets:
(1) a prefix-level pseudo-MOS computed on the truncated signal $x_{1:t}$, and 
(2) a full-utterance target computed on the complete signal $x_{1:T}$.
The full-utterance target includes all metrics defined in the original ARECHO metric set, while the prefix-level pseudo-MOS is obtained by applying the same reference-free estimators to the truncated signal $x_{1:t}$.
For complete utterances ($t = T$), the prefix-level and full-utterance targets coincide.

Prefix expansion yields 583,983 samples ($\approx$3.4$\times$ increase), which are split 80/20 into 467,657 training and 116,326 validation instances.

We note that this expansion introduces \emph{new supervision signals} at distinct truncation points rather than repeating full-utterance labels; the metric-specific pattern of gains (see Section~\ref{sec:results}) contradicts a pure data-volume explanation.
The Overall Dev split (8,700 utterances) yields 34,726 prefix instances after expansion; we evaluate on this split to ensure direct comparability with the ARECHO checkpoint.

\subsection{Supervision and Metrics}

We supervise prefix-level prediction using UTMOS, PLCMOS, and NISQA as reference-free pseudo-MOS targets. 
Full-utterance supervision uses the complete metric set defined in ARECHO (Section~\ref{sec:method}).
Since ARECHO uses identical supervision for full utterances, the comparison isolates the architectural contribution. All metric values are discretized as described in Section~\ref{sec:training_objective}.
We report Mean Absolute Error (MAE), Pearson Correlation (PCC), and Spearman Rank Correlation (SRCC), evaluated separately for chunk-level and full-utterance prediction.

\subsection{Hyperparameter Setup}

We initialize from a pretrained ARECHO model trained for full-utterance multi-metric prediction. The pretrained checkpoint is publicly available online.\footnote{\url{https://huggingface.co/espnet/arecho_scale_v0.1-large-decoder}} Specifically, WavLM-Large~\cite{Chen2022WavLM} is used as the acoustic frontend and frozen during finetuning. The audio encoder contains 4 Transformer layers, while the autoregressive decoder contains 12 Transformer layers with 8 attention heads and embedding dimension 256. The decoder vocabulary was expanded from 32926 to 65828 to include chunk-level and full-utterance metric tokens.

During training, we adopt AdamW optimizer with learning rate $4{\times}10^{-4}$ and a linear warmup over 50k steps. For loss computation, we adopt a label smoothing of 0.1. The batch size is set to 12 with a gradient accumulation of 2. As the model is initialized from a pre-trained model, we only set 15 epochs for training, when the final checkpoint is used for evaluation. During inference, we apply standard greedy decoding under the predefined resolution-aware token order (i.e., chunk-based metrics are first predicted).

\subsection{Baseline}

Our primary baseline is the pretrained ARECHO checkpoint~\cite{shi2025arecho}, 
which predicts multi-metric quality tokens from complete utterances using the 
same WavLM frontend and decoder architecture. ARECHO is trained exclusively 
with full-utterance supervision and has no mechanism for prefix-based or 
multi-resolution prediction. At evaluation, this checkpoint is applied directly 
to prefix inputs without adaptation, providing a controlled comparison that 
isolates the effect of ANCHOR's three additions: dual-resolution query tokens, 
chunk-first decoding order, and prefix-expanded training.

\section{Results}
\label{sec:results}

We organize results around three questions:
(1) Does chunk-first decoding improve prefix-based prediction?
(2) How quickly does full-utterance prediction converge as context grows?
(3) How do the two systems behave under controlled distortion?

\begin{figure}[t]
  \centering
  \includegraphics[width=0.8\linewidth]{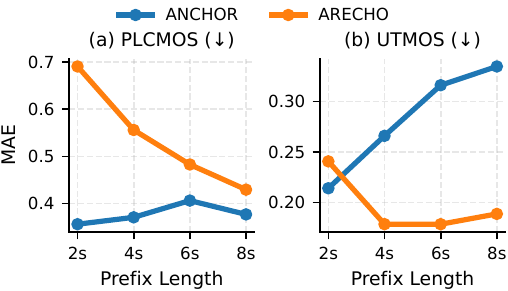}
  \caption{
  Chunk-level prediction MAE across prefix lengths.
  (a) PLCMOS: ANCHOR consistently outperforms ARECHO.
  (b) UTMOS: ANCHOR wins at 2\,s but a crossover emerges at longer prefixes.
  }
  \label{fig:tradeoff}
\vspace{-1.5em}
\end{figure}

\subsection{Chunk-Level: Local vs.\ Global Trade-off}

Figure~\ref{fig:tradeoff} compares ANCHOR and ARECHO on chunk-level (prefix-evaluated) prediction.

\textbf{ANCHOR dominates on local-sensitive metrics.}
For PLCMOS, ANCHOR reduces MAE by 48\% at 2\,s and maintains consistent gains across all prefix lengths (33\% at 4\,s, 16\% at 6\,s, 12\% at 8\,s). The chunk-first schedule, which commits to local estimates before global refinement, directly benefits metrics that capture short-time discontinuities.

\textbf{A structured crossover appears on global metrics.}
For UTMOS, ANCHOR improves at 2\,s (MAE: 0.241$\to$0.214, PCC: 0.935$\to$0.950), but ARECHO surpasses it beyond 2\,s. This is architecturally expected: as the decoder conditions on the preceding chunk tokens, its attention shifts toward local history at the expense of global integration. The trade-off is bounded and occurs where ARECHO already benefits from near-complete context. This asymmetry also serves as implicit ablation evidence: if the $3.4\times$ data expansion alone explained the gains, all metrics should improve uniformly, whereas the structured divergence matches the predicted signature of chunk-first ordering.

\subsection{Full-Utterance Convergence from Partial Context}

Table~\ref{tab:convergence} reports a different task from Section~5.1: how accurately ANCHOR predicts \emph{full-utterance} quality from each prefix length. ARECHO lacks prefix-to-full training and cannot perform this task.

\begin{table}[t]
\centering
\caption{ANCHOR full-utterance prediction from prefixes. MAE ($\downarrow$) and Pearson correlation LCC ($\uparrow$).}
\label{tab:convergence}
\vspace{0.3em}
\resizebox{\linewidth}{!}{
\begin{tabular}{lcccccccc}
\toprule
& \multicolumn{4}{c}{\textbf{MAE} ($\downarrow$)} & \multicolumn{4}{c}{\textbf{LCC} ($\uparrow$)} \\
\cmidrule(lr){2-5} \cmidrule(lr){6-9}
\textbf{Metric} & \textbf{2\,s} & \textbf{4\,s} & \textbf{6\,s} & \textbf{8\,s} & \textbf{2\,s} & \textbf{4\,s} & \textbf{6\,s} & \textbf{8\,s} \\
\midrule
PLCMOS     & 0.865 & 0.725 & 0.734 & 0.758 & 0.629 & 0.719 & 0.689 & 0.684 \\
UTMOS      & 0.236 & 0.183 & 0.184 & 0.176 & 0.934 & 0.959 & 0.963 & 0.968 \\
DNS        & 0.312 & 0.238 & 0.218 & 0.195 & 0.838 & 0.895 & 0.893 & 0.902 \\
NISQA-Noi  & 0.477 & 0.325 & 0.322 & 0.303 & 0.820 & 0.916 & 0.908 & 0.908 \\
NISQA-MOS  & 0.598 & 0.466 & 0.486 & 0.469 & 0.831 & 0.888 & 0.876 & 0.879 \\
\bottomrule
\end{tabular}
}
\vspace{-0.5em}
\end{table}

\textbf{Most perceptual evidence accumulates by 4\,s.}
Across all five metrics, the largest MAE reduction occurs between 2\,s and 4\,s. LCC patterns corroborate this: UTMOS correlation rises sharply in this window (0.934$\to$0.959), and DNS and NISQA-Noise show similar jumps. Beyond 6\,s, both MAE and LCC plateau.

\textbf{Convergence patterns are metric-dependent.}
UTMOS converges monotonically (LCC steadily increases to 0.968 at 8\,s). In contrast, PLCMOS exhibits non-monotonic behavior: MAE temporarily undershoots the 8\,s baseline at 4\,s, and LCC peaks at 4\,s (0.719) then declines. This over-correction is consistent with the chunk-first bias overweighting local cues. 

\begin{figure*}[!t]
  \centering
  \includegraphics[width=0.8\textwidth]{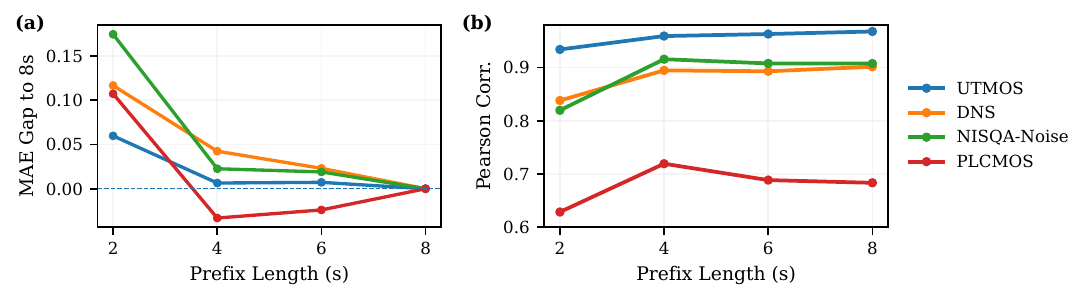}
\caption{
Prefix-to-full convergence under ANCHOR.
(a) MAE gap drops sharply between 2\,s and 4\,s, with diminishing returns thereafter.
(b) Pearson correlation stabilizes by 4–6\,s, indicating an effective context horizon.
}
\label{fig:convergence}
\end{figure*}

\textbf{An effective context horizon emerges at 4–6\,s.}
Figure~\ref{fig:convergence} visualizes these trends: the MAE gap drops sharply between 2\,s and 4\,s with diminishing returns beyond 6\,s, while Pearson correlation stabilizes by 4–6\,s across metrics. We interpret this 4–6\,s stabilization as an \emph{effective} context horizon: beyond roughly 4–6\,s, additional context yields little improvement in predicting full-utterance quality. The divergence across metrics (non-monotonic PLCMOS vs.\ monotonic UTMOS) indicates that this reflects the behavior of the metrics themselves rather than a purely architectural artifact.

\subsection{Controlled Distortion Stress Test}

We designed a tightly controlled test to isolate extrapolation behavior under a \emph{known, fixed} corruption, not to establish general robustness.
We injected at $t=1.5$\,s into 100 dev-set utterances: (i) a 100\,ms noise burst at 5\,dB SNR, and (ii) a 200\,ms packet drop (silence). Predictions on prefixes (2–8\,s) were compared against clean full-utterance ground truth (Table~\ref{tab:distortion_bias}).

\begin{table}[t]
\centering
\caption{Mean prediction bias under controlled distortion (100 utterances).
Positive = optimistic relative to clean full-utterance quality.}
\label{tab:distortion_bias}
\vspace{0.3em}
\resizebox{0.95\linewidth}{!}{
\begin{tabular}{lcc}
\toprule
\textbf{Metric} 
& \textbf{Mean Bias (ANCHOR)} 
& \textbf{Mean Bias (ARECHO)} \\
\midrule
PLCMOS        & $+0.257$ & $-0.140$ \\
SI-SNR        & $+1.075$ & $-2.427$ \\
SDR           & $+1.480$ & $-1.754$ \\
UTMOS         & $-0.104$ & $-0.156$ \\
NISQA-dist    & $-0.370$ & $-0.132$ \\
\bottomrule
\end{tabular}
}
\vspace{-0.5em}
\end{table}

\textbf{Signal-domain metrics show divergent extrapolation.}
ANCHOR exhibits consistent positive bias on SI-SNR and SDR, while ARECHO shows substantial negative bias, indicating fundamentally different behavior when extrapolating from corruption-containing prefixes.

\textbf{Perceptual predictors are more aligned but still differ.}
Both systems show mild negative bias on UTMOS and DNS, but ANCHOR is consistently less pessimistic. For NISQA distortion, ANCHOR is \emph{more} pessimistic, reflecting greater sensitivity to structural corruption.

\textbf{Bias direction is metric-dependent, not system-dependent.}
Across 13 evaluated metrics, ANCHOR is less pessimistic in 8 and positively biased in 6. The largest divergence occurs in signal-domain measures, confirming that prefix-based prediction under distortion is governed by what each metric captures rather than a uniform system-level effect.

\section{Conclusion}

We introduced ANCHOR, a unified autoregressive framework
for joint chunk-level and full-utterance speech quality modeling.
By enforcing a resolution-aware decoding hierarchy, we mitigate supervision conflict between local and global objectives.
Experiments show that ANCHOR substantially improves packet-loss–sensitive perceptual modeling under short-context conditions, while revealing a structured trade-off with global MOS prediction.
Prefix-to-full convergence analysis further suggests the presence of an effective context horizon, where approximately 4–6 seconds of speech provide sufficient information for stable approximation of full-utterance quality (within the 2–8\,s range evaluated).
A controlled distortion stress test further confirms that extrapolation bias is metric-dependent rather than system-dependent, with ANCHOR maintaining more stable behavior under localized corruption.
We note that ANCHOR currently relies on a non-causal frontend and therefore does not constitute a fully streaming system.
Nevertheless, it establishes a principled framework for prefix-based quality assessment, prioritizing local perceptual robustness to support future low-latency deployment scenarios. We acknowledge that formal component-wise ablations (e.g., interleaved vs.\ chunk-first ordering) were not included; however, the metric-specific pattern of gains in Section~\ref{sec:results} provides indirect evidence that the decoding hierarchy, rather than data expansion alone, drives the observed behavior.

\section{Acknowledgments}
This work used the Bridges-2 system at PSC and the Delta system at NCSA
through allocations CIS210014 and IRI120008P from the Advanced
Cyberinfrastructure Coordination Ecosystem: Services \& Support (ACCESS)
program~\cite{xsede,nystrom2015bridges,boerner2023access}, which is supported by U.S.
National Science Foundation grants \#2138259, \#2138286, \#2138307,
\#2137603, and \#2138296.

\section{Use of Generative AI Disclosure}
Generative AI tools were used to assist with code implementation and debugging, and for minor language editing of the manuscript. All experimental design, analysis, and scientific claims are the authors' own, and the authors take full responsibility for the work and its content.

\bibliographystyle{IEEEtran}
\bibliography{mybib}

\end{document}